\documentclass{eds2019}
\usepackage{amsmath,amssymb}
\usepackage{wrapfig}
\def\Journal#1#2#3#4{{#1} {\bf #2}, #3 (#4)}

\def\NPB{{\em Nucl. Phys.} B}
\def\PLB{{\em Phys. Lett.}  B}
\def\PRL{\em Phys. Rev. Lett.}
\def\PRD{{\em Phys. Rev.} D}

\def\ZPC{{\em Z. Phys.} C}
\def\EPL{\em EPL (Europhysics Letters)}

\def\be{\begin{equation}}
\def\ee{\end{equation}}
\def\bea{\begin{eqnarray}}
\def\eea{\end{eqnarray}}

\begin{document}
\vspace*{4cm}
\title{DIFFRACTIVE DISSOCIATION IN ONIUM - NUCLEUS SCATTERING FROM A PARTONIC PICTURE}

\author{ D. LE ANH}

\address{CPHT, CNRS, Ecole polytechnique, IP Paris, F-91128 Palaiseau, France}

\maketitle\abstracts{
We present a partonic picture for diffractive onium-nucleus scattering in the large-number-of-color limit from which the distribution of rapidity gaps in a certain kinematic region can be deduced. This picture allows us to draw a parallel between diffractive dissociation and the genealogy of partonic evolution, the latter being essentially similar to a branching-diffusion process in which the branching is the parton splitting, and the diffusion occurs in the transverse momenta of the partons. In particular, we show that the rapidity gap distribution corresponds to the distribution of the splitting rapidity of the last common ancestor of the partons whose transverse momenta are smaller than the nuclear saturation scale, when the scattering process is viewed in the restframe of the nucleus. Numerical calculations are also implemented to support the analytical predictions.}

\section{Introduction} \label{sec:intro}

Despite the fact that a hadron is a weakly bound system of partons in the context of high-energy scattering, there was a notable observation from the electron-proton collision at DESY - HERA that, in a proportion of about 10\% of the events, the scattered proton remains untouched~\cite{ZEUS,H1}. This phenomenon is defined to be hard diffraction and characterised by a large angular separation between the outgoing proton and the particle produced closest to the direction of the outgoing proton in the final state, which is corresponding to a large Lorentz-invariant rapidity gap $y_0$. Diffractive dissociation is also expected to happen in high-energy electron-nucleus collision at a future electron-ion collider (EIC)~\cite{EIC}.\\
In the work reported here, we focus on diffractive dissociation in the scattering of an onium (a color-singlet quark-antiquark pair) off a nucleus, which is directly related to the electron-nucleus scattering in small-$x$ limit (see Sec.~\ref{sec:kovchegov.levin} below) . In particular, we put an effort to find the event-by-event rapidity gap distribution and try to relate diffraction with ancestry problems in branching-diffusion processes. This paper summarises discussions and results from papers published previously \cite{mueller.munier.PRD.2018,mueller.munier.PRL.2018} (Sec. 3), and provides further original numerical results on the ancestry problem in comparison to diffraction (Sec. 4).

\section{Diffraction in the Kovchegov - Levin formalism} \label{sec:kovchegov.levin}

\begin{wrapfigure}{r}{0.58\textwidth} 
    \centering
    \includegraphics[width=0.54\textwidth]{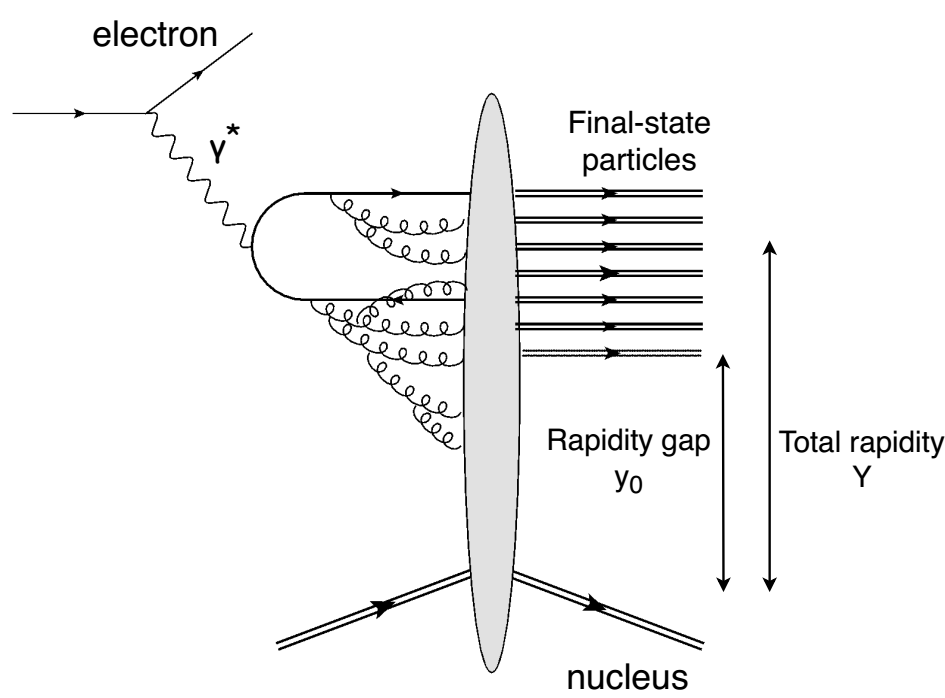}
    \caption{A diagramatic illustration of the diffractive scaterring. The virtual photon fluctuates to an onium, which, in turn, develops to a gluon shower before interaction with the nucleus. Only gluons of rapidity larger than $y_0$ (viewed from the nucleus) can be hadronized to become final-state particles while other gluons recombines back to the original onium, leaving a rapidity gap $y_0$.}
    \label{fig:diagram.illustr}
\end{wrapfigure}

The small-$x$ description of deep-inelastic scattering allows us to treat the virtual photon, which mediates the interaction between the electron and the nucleus, as a colorless $\rm q\bar{q}$ dipole (onium) in some appropriate reference frames (see Fig.~\ref{fig:diagram.illustr}). The S-matrix element $S(r,y)$ describing the scattering of an onium of size $r$ off a nucleus at the rapidity $y$ in QCD is governed by the so-called Balitsky-Kovchegov (BK) evolution equation~\cite{balitsky.1996,kovchegov.1999}, which is written in the framework of the color dipole model~\cite{mueller.1994} as
\begin{equation}
	\partial_{y} S(r,y) = \frac{\bar{\alpha}}{2\pi}\int d^2r^{\prime} \frac{r^2}{r^{\prime 2}(r-r^{\prime})^2}\left[S(r^{\prime},y)S(r-r^{\prime},y)-S(r,y)\right],
	\label{eq:balisky.kovchegov}
\end{equation} 
where the notation $\bar{\alpha} \equiv \alpha_sN_c/\pi$ is used ($N_c$ is the number of colors, and $\alpha_s$ is the strong coupling constant). The initial condition (at $y=0$) could be given by the McLerran-Venugopalan (MV) profile~\cite{mclerran.venugopalan.1994.1,mclerran.venugopalan.1994.2}:
\begin{equation}
	S_{MV}(r) = \exp{\left[-\frac{r^2Q_{MV}^2}{4}\ln{\left(e+\frac{4}{r^2\Lambda_{QCD}^2}\right)}\right]},
	\label{eq:MV}
\end{equation} 
which characterises the interaction between a bare onium of size $r$ with the nucleus. $Q_{MV}$ is the saturation momentum scale of the nucleus such that $1/Q_{MV}$ could be interpreted as the onium size above which the probability of scattering is of order unity.\\
As shown by Kovchegov and Levin~\cite{kovchegov.levin.2000,kovchegov.levin.2012}, the diffractive dissociation of the onium due to the interaction with the nucleus can be described by an auxiliary function $S_D(r,y,y_0)$, which corresponds to the probability of observing a gap larger than $y_0$ and obeys the BK equation:
\begin{equation}
	\partial_{y} S_D(r,y,y_0) = \frac{\bar{\alpha}}{2\pi}\int d^2r^{\prime} \frac{r^2}{r^{\prime 2}(r-r^{\prime})^2}\left[S_D(r^{\prime},y,y_0)S_D(r-r^{\prime},y,y_0)-S_D(r,y,y_0)\right],
	\label{eq:kovchegov.levin}
\end{equation} 
with the initial condition given by
\begin{equation}
	\label{eq:init.KL}
	S_D(r,y=y_0,y_0) = S^2(r,y_0).
\end{equation}
The diffractive cross-section can then be derived from $S_D$ by taking a $y_0$ - derivative, 
\begin{equation}
	\frac{d\sigma_D}{dy_0} = -\frac{\partial }{\partial y_0}S_D(r,y,y_0).
\end{equation}
Equations (\ref{eq:balisky.kovchegov}) and (\ref{eq:kovchegov.levin}), equiped with their initial conditions (\ref{eq:MV}) and (\ref{eq:init.KL}), are known as the Kovchegov-Levin (KL) system of equations for the diffractive onium-nucleus scattering~\footnote{In the original paper of Kovchegov and Levin~\cite{kovchegov.levin.2000}, instead of $S_D$, they derived an equation for $N_D = 1-2S+S_D$}. Due to the sophisticated structure of this system, the formalism is not helpful in understanding the analytic behavior of the rapidity gap distribution, which requires other approaches to get in on the act.

\section{Diffractive onium - nucleus scattering from a partonic picture}
\label{sec:picture}

\subsection{Rapidity gap distribution}

Let us start from the total cross-section and describe it in the frame in which the nucleus is at rest while the onium is boosted to the total rapidity of the scattering, which is hereafter denoted by $Y$. Boosting the onium to a higher rapidity gives possibility for additional soft gluons to be emitted. In the large-$N_c$ limit, a one-gluon emission by the onium can be viewed as a splitting of a dipole into two dipoles, in general, of different sizes. This splitting, which may be repeated for subsequent dipoles for further boost, makes the QCD evolution of the onium appear as a dipole branching-diffusion process. At the time of interaction, the onium interacts with the nucleus not as a bare dipole, but as a set of dipoles whose detail varies from event to event. For a scattering to occur with a significant probability, it is required to have at least one dipole in the onium Fock state at rapidity $Y$ whose size is larger that $1/Q_{MV}$. \\
Back to the diffractive onium-nucleus scattering, it is also beneficial to adopt the picture of event-by-event large-dipole fluctuations for the dipole evolution. Instead of being in the rest frame of the nucleus, a diffractive event is now modeled in a frame in which the nucleus is boosted to a rapidity $y_0$, leaving the remaining rapidity $\tilde{y}_0 = Y-y_0$ to the onium. Unlike the onium, the nucleus is a dense system of gluons whose evolution in rapidity is deterministic, and it is now characterised by the evolved saturation momentum $Q_{s}(y_0)$ defined by
\begin{equation}
	\label{eq:sat.scale}
	Q_s^2(y_0) = Q_{MV}^2\frac{e^{\bar{\alpha} y_0 \chi^{\prime}(\gamma_0)}}{(\bar{\alpha}y_0)^{3/(2\gamma_0)}},
\end{equation}
where $\chi(\gamma)=2\psi(1)-\psi(\gamma)-\psi(1-\gamma)$ is the eigenvalue of the kernel of the linearised BK equation (\ref{eq:balisky.kovchegov}) around $S\sim 1$ corresponding to the eigenfunction $1-S=r^{2\gamma}$. $\gamma_0$ is the solution of the equation $\chi^{\prime}(\gamma_0)=\chi(\gamma_0)/\gamma_0$ ($\gamma_0\approx 0.63$). For the onium, the stochastic nature of the evolution results in the fact that there may be some events in which the onium Fock state at the rapidity $\tilde{y}_0$ contains at least one exceptionally large dipole whose size is beyond $1/Q_s(y_0)$. For such realization of the onium state, the scattering with the nucleus in this particular frame has unit probability, and consequently, the probability of having an elastic scattering of this system with the nucleus is of order unity. Therefore, that generates a rapidity gap $y_0$, and the rapidity gap distribution $d\sigma_{D}/dy_0$ is then corresponding to the probability to have that type of events. Using the travelling wave solution of the BK equation~\cite{munier.peschanski.2004}, the argument above allows a straightforward derivation of the rapidity gap distribution, which is given by
\begin{equation}
\label{eq:gap.distrib}
	\frac{1}{\sigma_{tot}}\frac{d\sigma_D}{dy_0} = c_D \left[\frac{\bar{\alpha}Y}{\bar{\alpha}y_0(\bar{\alpha}Y-\bar{\alpha}y_0)}\right]^{3/2}.
\end{equation} 
The diffractive cross-section $d\sigma_{D}/dy_0$ normalised to the total cross-section $\sigma_{tot}$ in the equation (\ref{eq:gap.distrib}) gives a valid $y_0$-dependence for a fixed asymptotic large rapidity $Y$ in the so-called scaling region defined by
\begin{equation}
	\label{eq:scaling.region}
	1 \ll \left|\ln\left[r^2Q_s^2(Y)\right] \right| \ll \sqrt{\bar{\alpha}Y\chi^{\prime\prime}(\gamma_0)}.
\end{equation}
A more detailed calculation would show an additional Gaussian suppression factor of the form $\exp\left[-\frac{\ln^2\left[r^2Q_s^2(Y)\right]}{2\chi^{\prime\prime}(\gamma_0)\bar{\alpha}(Y-y_0)} \right]$ in the equation (\ref{eq:gap.distrib}), which tends to $1$ when moving deeply inside the scaling region (\ref{eq:scaling.region}). The understanding of the rapidity gap distribution from this elegant formula is, however, not complete since the constant $c_D$ cannot be determined from this approach.
\subsection{Diffraction and ancestry problem for dipole evolution}

The picture for the diffractive onium-nucleus scattering induces another interesting problem related to the genealogy of the dipole evolution. Returning to the nucleus rest frame, one now considers all dipoles of size larger that $1/Q_{MV}$ and trace the evolution of the onium back to the rapidity $y_0$ (viewed from the nucleus) where the last common ancestor of these extreme particles splits. The distribution of this branching rapidity, $G(r,y,y_0)$, is governed by the following exact ancestry equation~\cite{le.2018,le.munier.2018}:
\begin{multline}
	\label{eq:ancestry.eq}
	\partial_y G(r,y,y_0) = \frac{\bar{\alpha}}{2\pi} \int d^2r^{\prime} \frac{r^2}{r^{\prime 2}(r-r^{\prime})^2} \left[G(r^{\prime},y,y_0)S(r-r^{\prime},y)\right.\\
	\left. + \ G(r-r^{\prime},y,y_0)S(r^{\prime},y)- G(r,y,y_0) \right],
\end{multline}
with the initial condition
\begin{equation}
	\label{eq:init.ancestry}
	G(r,y=y_0,y_0) = \frac{\bar{\alpha}}{2\pi} \int d^2r^{\prime} \frac{r^2}{r^{\prime 2}(r-r^{\prime})^2} \left[ 1 - S(r^{\prime},y_0)\right]\left[ 1 - S(r-r^{\prime},y_0)\right]. 
\end{equation}
The function $S(r,y)$ that appears in this equation is again the S-matrix element, which obeys the BK equation (\ref{eq:balisky.kovchegov}). Although it is intractable to solve the ancestry equation for $G(r,y,y_0)$ analytically, its behavior could be understood by adopting an appealing relation between the diffraction and the ancestry problems. Indeed, if the initial onium size is chosen so that it is far away from the saturation boundary of the nucleus, as indicated by the leftmost inequality in the condition (\ref{eq:scaling.region}), the last common ancestor of the extreme particles is due to a large fluctuation at $y_0$ which generates a dipole of size beyond $1/Q_s(y_0)$ with a high possibility. In this sense, the distribution of the branching rapidity of the last common ancestor is tantamount to the distribution of the rapidity gap up to an overall constant, in the scaling region. \\
As discussed previously, the QCD evolution of the onium is a dipole branching-diffusion process, which turns out to be similar to a branching Brownian motion (BBM). Consider an one-dimensional BBM along the $x$-axis in time $t$, with the diffusion coefficient $D$ and the branching rate $r$. The mean density of particles $u$ obey the equation $\partial_t u = \hat{\eta} u$, where $\hat{\eta} = D\partial_x^2+r$ is the branching-diffusion kernel. The kernel $\hat{\eta}$ admits the eigenfunction $e^{-\beta x}$, which corresponds to the eigenvalue $\eta (\beta)=D\beta^2+r$. At the final time $t=T$, let us pick $n$ ($n=2,3,\ldots$) leftmost (or rightmost) particles in each realization and search for the spliting time $t_0$ of their last common ancestor. Derrida and Mottishaw~\cite{derrida.mottishaw.2016} recently found that, the distribution $p(t_0|T)$ of that splitting time is given by
\begin{equation}
	\label{eq:derrida.mottishaw}
	p(t_0|T) = \frac{1}{\beta_0}\frac{1}{\sqrt{2\pi\eta^{\prime\prime}(\beta_0)}} \left[\frac{T}{t_0(T-t_0)}\right]^{3/2},
\end{equation}
where $\beta_0$ solves the equation $\eta(\beta_0)=\beta_0\eta^{\prime}(\beta_0)$. This problem is not equivalent, but in the similar fashion to the ancestry problem in the context of dipole evolution. In fact, this distribution is precisely the same as the rapidity gap distribution in the equation (\ref{eq:gap.distrib}) up to the substitutions $\bar{\alpha}Y \leftrightarrow T$ and $\bar{\alpha}y_0 \leftrightarrow t_0$, and to an overall constant. Another appealing feature of the ancestry distribution in the BBM context is that, the overall normalization factor is definitely determined. The translation of this factor from the BBM to the dipole branching-diffusion can be done by the effective replacements $\eta(\beta)\leftrightarrow\chi(\gamma)$ and $\beta_0\leftrightarrow\gamma_0$, so that the overall constant in the equation (\ref{eq:derrida.mottishaw}) for the latter case should read 
\begin{equation}
	\label{eq:conjecture.constant}
	c = \frac{1}{\gamma_0}\frac{1}{\sqrt{2\pi\chi^{\prime\prime}(\gamma_0)}}.
\end{equation}
One may conjecture that this constant could be the unknown constant $c_D$ in the equation (\ref{eq:gap.distrib}) by the diffraction-genealogy parallel. We shall examine this analogy numerically in the next section.

\section{Numerical results}
\label{sec:numerics}

Although it is impossible to solve analytically the KL and ancestry systems of equations, the numeric integration of these systems could be implemented quite straightforwardly. We are going to test the validity of the distribution (\ref{eq:gap.distrib}), together with the diffraction-genealogy analogy for large values of the total rapidity $Y$.
\begin{figure}[h!]
	\centering
	\includegraphics[width=0.62\textwidth]{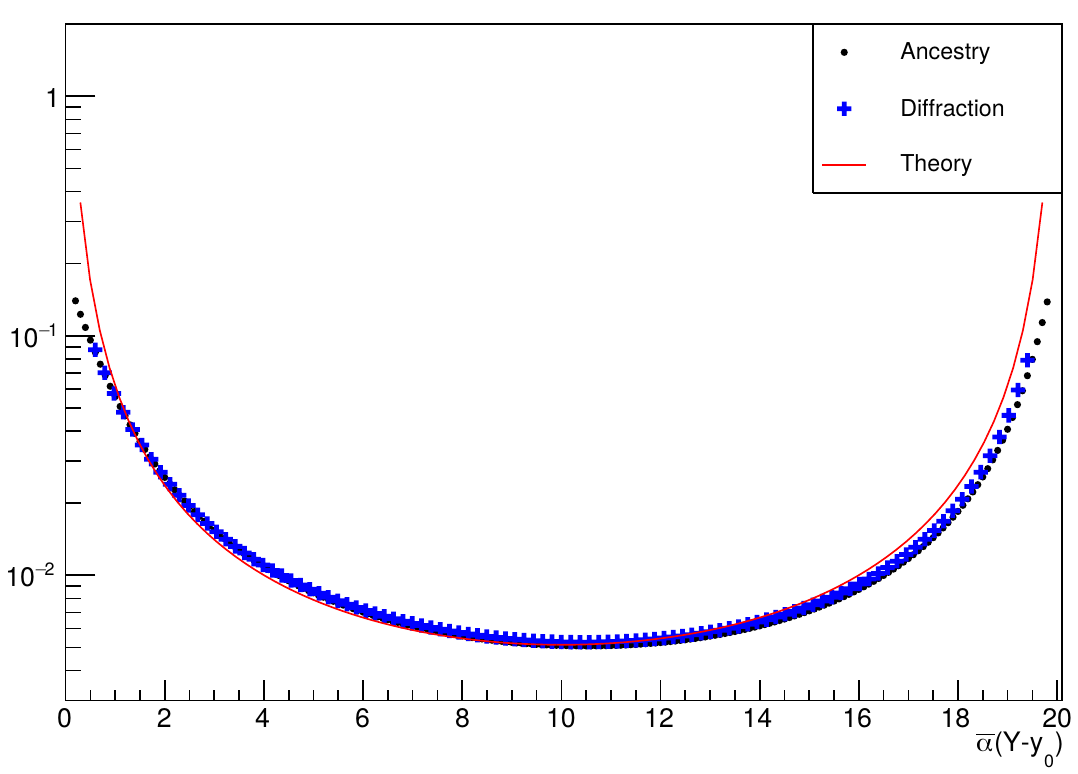}
	\caption{Numerical solutions for $\frac{1}{\sigma_{tot}}\frac{d\sigma_D}{dy_0}$ (``Diffraction") and $\frac{G}{1-S}$ (``Ancestry") with the total rapidity $\bar{\alpha}Y = 20$ and the onium size $rQ_{MV}= 4.0\times 10^{-21}$ as functions of $\bar{\alpha}(Y-y_0)$. The red continuous curve is the analytical prediction (\ref{eq:gap.distrib}) (``Theory") with $c_D = \frac{1}{\sqrt{2\pi\chi^{\prime\prime}(\gamma_0)}}$. This figure is adapted from Ref. [15].} 
	\label{fig:diffrac.ancestry.theory}
\end{figure}

\begin{figure}[h!]
\centering
	\begin{minipage}[t]{0.488\textwidth}
	\includegraphics[width=1.05\textwidth]{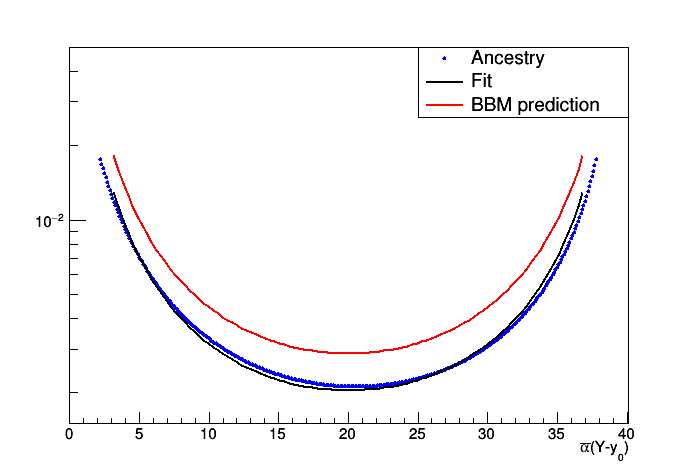}
	\caption{Numerical solution for $\frac{G}{1-S}$ (the blue dotted line) with $\bar{\alpha}Y = 40$ and $rQ_{MV}= 8.56\times 10^{-43}$ as a function of $\bar{\alpha}(Y-y_0)$ as compared to the BBM prediction (the red continuous line). The overall constant from the fit (the black continuous line) is $c_{\rm fit} \approx \frac{1.13}{\sqrt{2\pi\chi^{\prime\prime}(\gamma_0)}}$.}
	\label{fig:ancestry.bbm.fit}
	\end{minipage}
	\hfill
	\begin{minipage}[t]{0.488\textwidth}
	\includegraphics[width=1.05\textwidth]{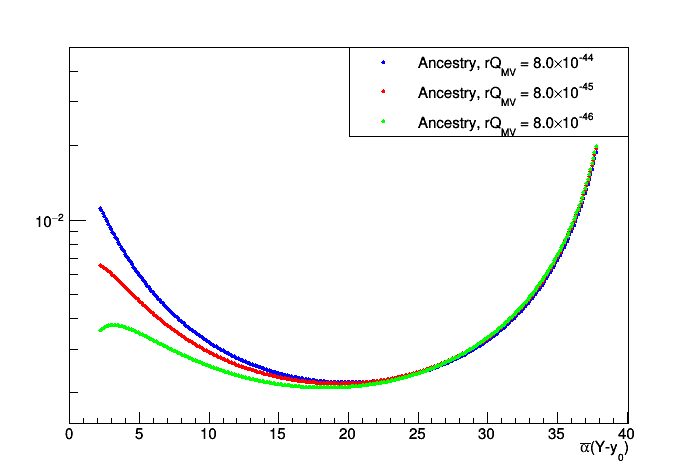}
	\caption{Numerical solutions for $\frac{G}{1-S}$ (Ancestry) with $\bar{\alpha}Y = 40$ as functions of $\bar{\alpha}(Y-y_0)$ for the onium size $rQ_{MV}$ approaching the rightmost boundary of the scaling region (\ref{eq:scaling.region}).}
	\label{fig:ancestry.scaling.boundary}
	\end{minipage}
\end{figure}

\noindent Numerics for $\bar{\alpha}Y=20$ and for the onium size inside the scaling region (\ref{eq:scaling.region}) exhibits a good agreement between the normalised distributions of the rapidity gap and the ancestor's splitting rapidity, as shown in Fig.~\ref{fig:diffrac.ancestry.theory}. They all match the analytic behavior (\ref{eq:gap.distrib}) fairly well as expected. Switching to a higher value of the total rapidity, $\bar{\alpha}Y=40$, one obtains a better fit to the analytic distribution (Fig.~\ref{fig:ancestry.bbm.fit}), with the fitted overall constant closer to the conjectured value (\ref{eq:conjecture.constant}) from the BBM distribution, as compared to the previous case with $\bar{\alpha}Y=20$. As moving towards the rightmost boundary of the scaling region (\ref{eq:scaling.region}), the Gaussian suppression, as mentioned in the previous section, becomes significant (Fig.~\ref{fig:ancestry.scaling.boundary}). This observation further verifies the reliability of the present approach. 

\section*{Acknowledgments}

This  research  was  supported in part by the Agence Nationale de la Recherche under the project \# ANR-16-CE31-0019.

\end{document}